\documentstyle[12pt]{article}
\def\be{\begin{equation}}
\def\ee{\end{equation}}
\def\barr{\begin{array}}
\def\earr{\end{array}}

\def\etal{ {\it et al.} }

\def\rp{$R_p \hspace{-1em}/\;\:$}
\begin{document}
\setcounter{page}{0}
\renewcommand{\thefootnote}{\fnsymbol{footnote}}
\thispagestyle{empty}
\vspace*{-1in}
\begin{flushright}
CERN-TH/95--89 \\[2ex]
{\large \bf hep-ph/9504314} \\
\end{flushright}
\vskip 45pt
\begin{center}
{\Large \bf \boldmath New LEP bounds on $B$-violating scalar
couplings: \\[1.5ex]
$R$-parity violating supersymmetry or diquarks}

\vspace{11mm}
\bf
 Gautam Bhattacharyya\footnote{ gautam@cernvm.cern.ch},
 Debajyoti Choudhury\footnote{debchou@surya11.cern.ch}
 and {\bf K. Sridhar\footnote{sridhar@vxcern.cern.ch} }\\

\vspace{13pt}
 {\bf Theory Division, CERN, \\ CH--1211 Gen\`eve 23, Switzerland.}

\vspace{50pt}
{\bf ABSTRACT}
\end{center}

\begin{quotation}
We  use the precision electroweak data at LEP to place bounds
on  $B$-violating Yukawa couplings, two theoretically appealing examples
being provided by $R$-parity--violating
supersymmetry and diquarks.
The couplings  involving  the third generation
quarks are most severely constrained. These bounds are complementary
to those obtained from low-energy processes.

\end{quotation}

\vspace{170pt}
\noindent
\begin{flushleft}
CERN-TH/95--89\\
April 1995\\
\end{flushleft}

\vfill
\newpage
\setcounter{footnote}{0}
\renewcommand{\thefootnote}{\arabic{footnote}}

\setcounter{page}{1}
\pagestyle{plain}
\advance \parskip by 10pt

One of the goals of the precision measurements  on the $Z$ peak at LEP
is to probe scenarios  beyond the Standard Model (SM)
predicting the existence of new particles coupling to the $Z$.
The experimental sensitivity is enhanced if these particles have
tree-level couplings to the SM fermions as well. Particular
examples are provided by $R$-violating supersymmetry, as also
by  diquarks, the
 existence of which can be motivated  in the context of both Grand Unified
Theories as well as composite models\cite{hew}
\footnote{It may be noted that these theories  predict the
existence of leptoquarks as well. However, the presence of certain
types of leptoquarks in association with diquarks would lead to
rapid proton decay. In this letter we assume that leptoquark
couplings, if existing, are highly suppressed.}. Currently, there exist
phenomenological bounds --- from low-energy data ---
on only some of these couplings. We aim to use the LEP data on $Z$
partial widths to impose complementary bounds.

A diquark can, in general, be defined as an elementary integral-spin
particle with a baryon number $|B| = 2/3$ and lepton number $|L| = 0$
and  coupling to a pair of quarks.
Assuming the SM fermion content, the diquarks may transform as ${\bf 3}$
or ${\bf \bar{6}}$ under $SU(3)_C$;
as triplet or singlet under $SU(2)_L$;
 and can have electric charges
$|Q_D| = 1/3, 2/3$ or 4/3. In this letter, we restrict ourselves to a
discussion of scalar diquarks only. The Yukawa interaction terms
  can then be parametrized as:
\be
{\cal L}_Y = h^{(A)}_{ij} \overline{q_i^c} P_{L,R} q_j \phi_A +
{\rm h.c.}\ ,
\label{lagdq}
\ee
where $i$ and $j$ denote the flavour generation indices, $A$
labels the diquark-type and $h^{(A)}_{ij}$ are the corresponding
Yukawa couplings. The relevant projection operator $P_{L,R}$ depends
upon the gauge transformation properties of the concerned fields.
All such possible contributions within the texture of
the SM  are listed in Table 1. The coupling of the diquark
to the gauge bosons is obviously determined by the quantum numbers.
In addition, it will have self-coupling terms, and also couplings
with the SM Higgs, but these are of no consequence to us.

\begin{table}[htb]
\vspace*{4ex}
$$
\barr{|| c | l | c | l ||}
\hline
 \multicolumn{1}{||l|}{\rm Scalar} &  &
         \multicolumn{1}{c|}{ SU(3)_c \otimes SU(2)_L \otimes U(1)_Y} &  \\
\multicolumn{1}{||l|}{\rm type} & \multicolumn{1}{c|}{\rm Coupling}
       & \multicolumn{1}{c|}{\rm transformation} &
         \multicolumn{1}{c||}{\rm Remarks} \\
\hline
&&&\\
\phi_1 & \overline{(Q_{Li})^c} Q_{Lj} \phi_1 & (\bar{6},3, -1/3)
       &  {\rm Generation \ symmetric} \\
\phi_2 & \overline{(Q_{Li})^c} Q_{Lj} \phi_2 & (3,3, -1/3)
       &  {\rm Generation \ antisymmetric} \\
&&&\\[-.5ex]
\phi_3 & \overline{(Q_{Li})^c} Q_{Lj} \phi_3 & (\bar{6},1, -1/3)
       &  {\rm Generation \ antisymmetric} \\
\phi_4 & \overline{(Q_{Li})^c} Q_{Lj} \phi_4 & (3,1, -1/3)
       &  {\rm Generation \ symmetric} \\
&&&\\[-.5ex]
\phi_5 & \overline{(u_{Ri})^c} u_{Rj} \phi_5 & (\bar{6},1, -4/3)
       &  {\rm Generation \ symmetric} \\
\phi_6 & \overline{(u_{Ri})^c} u_{Rj} \phi_6 & (3,1, -4/3)
       &  {\rm Generation \ antisymmetric} \\
&&&\\[-.5ex]
\phi_7 & \overline{(u_{Ri})^c} d_{Rj} \phi_7 & (\bar{6},1, -1/3)
       &  {\rm No \ symmetry \ property} \\
\phi_8 & \overline{(u_{Ri})^c} d_{Rj} \phi_8 & (3,1, -1/3)
       &  {\rm No \ symmetry \ property} \\
&&&\\[-.5ex]
\phi_9 & \overline{(d_{Ri})^c} d_{Rj} \phi_9 & (\bar{6},1, 2/3)
       &  {\rm Generation \ symmetric} \\
\phi_{10} & \overline{(d_{Ri})^c} d_{Rj} \phi_{10} & (3,1, 2/3)
       &  {\rm Generation \ antisymmetric} \\
\hline
\earr
$$
\caption{The possible diquark couplings within the SM quark content.}
   \label{quantum}
\end{table}

The above Yukawa interaction is of a rather general type, parts of
which are mimiced by the $B$-violating parts of the $R$-parity-violating
(\rp) Yukawa interaction in supersymmetric theories \cite{rpar,barbieri}.
Representable as
$R = (-1)^{3B + L + 2 S}$, where $B,L,S$ are the baryon number, lepton
number and the intrinsic spin of the field, respectively,
$R$ has a value of $+1$ for
all SM particles and $-1$ for all their superpartners.
The $B$-violating part of the \rp~superpotential is
\be
{\cal W} = \lambda''_{ijk} U^c_i D^c_j D^c_k \ ,
\label{superrp}
\ee
where $U_i^c$ and $D_j^c$ are the up- and down-type quark superfields
respectively. The couplings
$\lambda''_{ijk}$ are obviously
antisymmetric in the last two indices.
The above interaction can be rewritten in terms of the component
fields as:
\be
{\cal L}_{R_p \hspace{-0.5em}/\;\:}
= \lambda''_{ijk} \left(u^c_i d^c_j \tilde{d}^*_k +
u^c_i \tilde{d}^*_j d^c_k + \tilde{u}^*_i d^c_j d^c_k\right) + {\rm h.c.}
\label{lagrp}
\ee
It is quite apparent that eq.(\ref{lagrp}) replicates parts of
eq.(\ref{lagdq}) when the quark fields
in the latter are $SU(2)_L$-singlets.
To be precise, we have a situation where scalars of
both type $\phi_8$ ($\tilde{d}_i$) and $\phi_{10}$ ($\tilde{u}_i$)
exist simultaneously.

It may be noted that there are important cosmological constraints
\cite{cosm} on the $R$-parity--violating Yukawa interactions.
Requiring that GUT-scale baryogenesis does not get washed out
imposes $\lambda''\ll 10^{-7}$ generically, though these bounds are
model dependent and can be evaded \cite{dr}. Evidently, similar
considerations also hold for diquarks.

As for phenomenological bounds on the diquarks, very little exists.
The preferred arena for their
production is obviously a hadron collider.
However, the large QCD background to the signal for diquark
pair-production results in poor detectability. One might think that
the situation is slightly better for \rp\ interactions,
as there are other
channels through which the squarks might decay\footnote{Indeed,  the
CDF  experiment at the Tevatron has quoted lower bounds on the
squark mass of $\sim 150$ GeV, modulo certain assumptions about the
supersymmetric parameter space \protect\cite{tevatron}.}.
It must be remembered  though, that in
the presence of \rp\ couplings, the lightest supersymmetric partner is
no longer stable, and many of the bounds do not hold. Moreover, the
 detection of a squark {\em per se} is not a valid indicator of a
$\Delta B = 1$ \rp\ interaction.

Some bounds have been obtained from low-energy processes though. For
example, the data on neutral meson mixing or that on $CP$--violation
in the $K$--sector can tightly constrain certain products of such
couplings \cite{barbieri}. As for individual bounds, strong ones
exist only for certain cases where both quarks are light.
These are derived from the non-observance of
 neutron--antineutron oscillations \cite{biswa,sher} and double
nucleon decay into kaons of identical strangeness \cite{sher}.

In this paper, we consider the effects of the Yukawa interactions
of the diquarks (and also of the squarks through their $B$-violating
Yukawa interactions) on the process $Z\rightarrow q \bar{q}$,
where $q$ is  a quark.
These interactions proceed through triangle and self-energy diagrams
with $Z, q$ and $\bar{q}$ as external legs. In each triangle and
self-energy diagram the internal lines are the diquark(s) and the
quark(s) which couple(s) to the diquark.
A typical set of triangle and self-energy diagrams with
a generic diquark (quark) $\phi$ ($Q$) are shown in Fig. 1.
The magnitude of the  contribution grows as the mass of
the internal quark in this case. Hence we focus our attention
on the terms involving the top quark.

The tree-level $Z$ couplings to the left- and right-handed fermions
can be parametrized as
\begin{equation}
M_\mu^{\rm tree} = e  \bar{q}(p^\prime) \gamma_\mu
(a_L^q P_L + a_R^q P_R) q(p).
\end{equation}
where
\begin{eqnarray}
a_L^q & = & (t_3^q - Q_q s_W^2)/s_W c_W, \nonumber  \\
a_R^q & = & - Q_q s_W /c_W.
\end{eqnarray}
The $Z$ couplings to the charge-conjugated fermions ($q^c$)
are, therefore, given by
\begin{equation}
a_L^{q^c} = - a_R^q, ~~~~~~~~~~~~~~~~~ a_R^{q^c} = - a_L^q.
\end{equation}
We compute the  self-energy and the vertex correction
diagrams in terms of the Passarino-Veltman $B$- and $C$-functions
\cite{pasvel}, corresponding to the two- and three-point
integrals\footnote{We use the numerical codes developed by
Mukhopadhyaya and Raychaudhuri in the context of \protect\cite{amit}.}.
Assuming for the sake of presentation that in Fig. 1 the external
fermions lines are {\em left-handed}, it is easy to see that
only the couplings of the left-handed fermions to the $Z$ are
modified, so that the new contribution to the amplitude is
given by
\begin{equation}
\label{e5}
M_{\mu}^{(i)} = {e h^2 N_c \over 16 \pi^2}
\; \bar q (p^{\prime}) \; \gamma_{\mu} \;A_i \;P_L \;q(p) ,
\end{equation}
where $i = 1,2,3$, $h$ is a generic Yukawa coupling, and
$N_c$ is the colour factor which is 3 (2) when
$\phi$ belongs to a ${\bf\bar{6}}$ (${\bf 3}$) of $SU(3)_C$.
The $A_i$'s are given by,
\begin{eqnarray}
\label{e6}
A_1 & = & a_L^Q m_t^2 C_0 -
a_R^Q \lbrace m_Z^2 (C_{22}-C_{23}) + (d-2)C_{24} \rbrace, \nonumber \\
A_2 &=& \displaystyle -2
\frac{t_3^\phi - Q_\phi s_W^2}{s_W c_W} {\tilde C}_{24},
\nonumber \\[1ex]
A_3 &=&  a_L^q B_1.
\end{eqnarray}
Here $A_{1,2}$ denote the contributions from the first and the
second triangle diagrams respectively, and the contributions of the two
self-energy diagrams are jointly denoted by $A_3$.  In $A_2$, we
use $\tilde{C}_{24}$ to distinguish it from the $C_{24}$ appearing
in $A_1$, as the structures of the propagators for the two triangle
diagrams are different. In the expression for $A_1$, $d$ denotes the
space-time dimension. Although the
individual diagrams are divergent, their sum is finite. The asymptotic
forms for $A_i$ can be found in refs. \cite{lepto}, which deal with
similar bounds on lepton number--violating Yukawa couplings.

To compare our results with the experimental numbers we use the following
observables:
\begin{enumerate}
\item
$ \displaystyle R_l = \Gamma_{\rm had} /\Gamma_l$,
which is stable under variation of the
 top-quark mass\footnote{We assume leptonic
universality.}. Recent measurements \cite{lep}
give $R_l^{\rm exp} = 20.795 \pm 0.04$ whereas the
SM prediction is $R_l^{\rm SM}
= 20.786$ for
a choice of $m_t = 175$ GeV, $m_H = 300$ GeV and $\alpha_S = 0.12$.

\item
$\displaystyle R_b = \Gamma_b / \Gamma_{\rm had}$,
which has a quadratic top mass dependence. From ref.\cite{lep},
$R_b^{\rm exp} = 0.2202 \pm 0.0020$ and $R_b^{\rm SM} = 0.2158$ for
the above choice of input parameters.

\item
$\displaystyle R_c = \Gamma_c / \Gamma_{\rm had} $,
which again has a quadratic top mass dependence. Ref.\cite{lep} quotes
$R_c^{\rm exp} = 0.1583 \pm 0.0098$ and $R_c^{\rm SM} = 0.172$ for
the same choice of input parameters as above.
\end{enumerate}

In Figs. 2--4, we plot respectively the deviations $\delta R_l$,
$\delta R_b$ and $\delta R_c$ caused by the presence of
$\phi_A \; (A = 2,4,6,8)$
as a function of the  scalar
mass $m_\phi$. In each case, we assume that
only  the corresponding
 Yukawa coupling is non-zero and equals unity.
 In Table 2 we show limits on the Yukawa
couplings of each scalar type for a common scalar mass of 100 GeV.

\begin{table}[ht]
\begin{center}
\bigskip
\begin{tabular}{||c||c|c|c||c|c|c||}
\hline
Coupling & \multicolumn{3}{| c ||}{Bounds from $R_l$}
         & \multicolumn{3}{| c ||}{Bounds from $R_b$} \\[1ex]
\cline{2-7}
& ($1 \sigma$) & ($2 \sigma$) & ($3 \sigma$)
& ($1 \sigma$) & ($2 \sigma$) & ($3 \sigma$) \\
\hline
&&&&&&\\[.2ex]
$h^{(1)}_{33}$ & 0.35 & 0.53 & 0.66 & - & - & 0.41 \\
$h^{(1)}_{13}$, $h^{(1)}_{23} $ &
  0.89 & 1.19 & 1.44 & - & - & 2.39 \\
\hline
&&&&&&\\[-.75ex]
$h^{(2)}_{13}$, $h^{(2)}_{23} $ &
  1.09 & 1.46 & 1.76 & - & - & 2.93 \\
\hline
&&&&&&\\[-.75ex]
$h^{(3)}_{13}$, $h^{(3)}_{23} $ &
  0.54 & 0.82 & 1.03 & 1.89 & 2.17 & 2.41 \\
\hline
&&&&&&\\[-.75ex]
$h^{(4)}_{33}$ &
  0.60 & 0.91 & 1.14 & - & - & 0.70 \\
$h^{(4)}_{13}$, $h^{(4)}_{23} $ &
  0.66 & 1.00 & 1.26 & 2.32 & 2.65 & 2.95 \\
\hline
&&&&&&\\[-.75ex]
$h^{(5)}_{13}$, $h^{(5)}_{23} $ &
  0.98 & 1.31 & 1.58 & - & - & 1.73 \\
\hline
&&&&&&\\[-.75ex]
$h^{(6)}_{13}$, $h^{(6)}_{23} $ &
  1.19 & 1.61 & 1.94 & - & - & 2.12 \\
\hline
&&&&&&\\[-.75ex]
$h^{(7)}_{33}$ &
  1.12 & 1.69 & 2.12 & - & - & 1.31 \\
$h^{(7)}_{31}$, $h^{(7)}_{32} $ &
  1.11 & 1.68 & 2.09 & 4.94 & 5.66 & 6.29 \\
\hline
&&&&&&\\[-.75ex]
$h^{(8)}_{33}$ &
  1.37 & 2.08 & 2.60 & - & - & 1.60 \\
$h^{(8)}_{31}$, $h^{(8)}_{32} $ &
  1.36 & 2.05 & 2.57 & 6.05 & 6.93 & 7.71 \\
\hline
\hline
&&&&&&\\[-.75ex]
$\lambda''_{313}$, $\lambda''_{323}$&
  0.97 & 1.46 & 1.83 & - & - & 1.89 \\
$\lambda''_{312}$ &
  0.96 & 1.45 & 1.82 & 4.28 & 4.90 & 5.45 \\
\hline
\end{tabular}
\caption[] {The  upper bounds on the diquark (and \rp) couplings for
$m_\phi$ = 100 GeV. Since $R_b^{\rm SM}$  is itself inconsistent with
$R_b^{\rm exp}$ at the $2 \sigma$ level, for most of the cases, $R_b$
does not give a bound at the $1 \sigma$ or $2 \sigma$ level.}
  \label{bounds}
\end{center}
\end{table}

We summarize our results below:
\begin{itemize}
\item
 The bounds obtained from $R_l$ are always better than those from
$R_b$, mainly because $R_l$ is measured with an accuracy of $0.2 \%$,
while $R_b$ is measured with an accuracy of not better than $1 \%$ at
this stage. Note, however, that many of the diquark couplings
lead to a negative contribution to $R_b$. Thus, with an improvement
in this determination, all such couplings can be constrained to a greater
extent.

\item
 We do not use the experimental numbers on $R_c$ to put any bounds, as
the experimental errors are relatively large. Notice, though, that
{\em all} of the diquark couplings lead to a {\em positive} contribution
to $R_c$. Thus, if the experimental value of $R_c$ continues to
stay below the SM prediction once the errors are reduced significantly,
this measurement will disfavour such scalars.

\item
 Only those contributions have been shown in figures which are
induced by $SU(3)_C$-triplet scalars. In cases where a ${\bf \bar{6} }$
scalar may couple as well, it is evident that its
contribution is similar to that of the ${\bf 3}$, except
for a colour enhancement factor of $3/2$.
The corresponding bounds on its Yukawa couplings
are thus stronger by a factor  of $\sqrt{3/2}$,
as can be seen from Table 2.

\item
 Scalars which couple to the top quark are constrained more
stringently than others because the largest
contributions\footnote{We find  that the next-to-leading terms
can be important in some cases, though.}  to
$\delta R_l$, $\delta R_b$ or  $\delta R_c$
are ${\cal O} (m^2_Q/m^2_\phi)$, where $Q$ represents the heavy quark
in the diagram.
Consequently, the bounds on $h^{(9)}_{ij} $ and $h^{(10)}_{ij}$
obtained by our method are an order of magnitude weaker than the
rest; hence
we do not include those  cases either in the figures or in Table 2.

\item
 Since the tree level predictions for $\Gamma_d$ and $\Gamma_s$
are nearly the same and also since $m_d, m_s \ll m_t, m_\phi$,
$h^{(A)}_{13} \simeq h^{(A)}_{23}$ for all $A$.

\item
 For $A =1$, $h_{33}$ is more strongly constrained than
$h_{13}$ or $h_{23}$. The reasons are twofold.
For one,  the presence of the
top  in both quark doublets serves to enhance the effect
for $h_{33}$. Furthermore, in the case of $h_{13}$ (and similarly
for $h_{23}$), while $\delta \Gamma_d < 0$, its effect is
negated to an extent by the fact that $\delta \Gamma_u > 0$.
This cancellation is obviously absent for $h_{33}$.

\item
 As is apparent from eq.(\ref{lagrp}), the \rp~ effects can be
realized by a linear combination of the contributions due to
scalars of the type
$\phi_8$ and $\phi_{10}$.  For $i = 3$  the dominance of the
former type is overwhelming due to the presence of the
heavy top quark in the loop\footnote{For the same reason, we may safely
ignore the $\tilde{t}_L$--$\tilde{t}_R$ mixing
in our computations.}.
In Table 2, we show the bounds on $\lambda''$-couplings
separately although we do not include them in figures.
That the bounds are a factor of $\approx \sqrt{2}$ stronger than
those on $h^{(8)}_{3i}$ can be understood from the fact that
 \rp couplings lead to a change in the $Z$ partial
width into two quark channels, unlike only one for the diquark, as
is evident from a comparison of eq.(\ref{lagdq}) and eq.(\ref{lagrp}).

\item
 The bounds on $\lambda''$-couplings obtained \cite{biswa,sher}
from the {\em assumption} of perturbative unification are $\sim$ 1.25,
which are at par with the phenomenological bounds we obtain.

\end{itemize}

In conclusion, we have obtained phenomenological bounds on $B$-violating
Yukawa couplings from LEP data on the $Z$ partial widths.
$\lambda''$-type Yukawa couplings in
\rp~ supersymmetric theories and the Yukawa couplings of the diquarks
with the standard quarks are two viable candidates of the above type.
Our method leads to significant bounds on those couplings which involve
the third generation quarks and hence
are complementary to those obtained from $n$--$\bar{n}$ oscillation or
other low-energy processes.
The limits we obtain will improve with the accumulation of
more data at LEP.

{\bf Acknowledgement} We would like to thank John Ellis for
useful discussions.

\begin{figure}[htb]
\vskip 9.5in\relax\noindent\hskip -1.5in\relax{\includegraphics{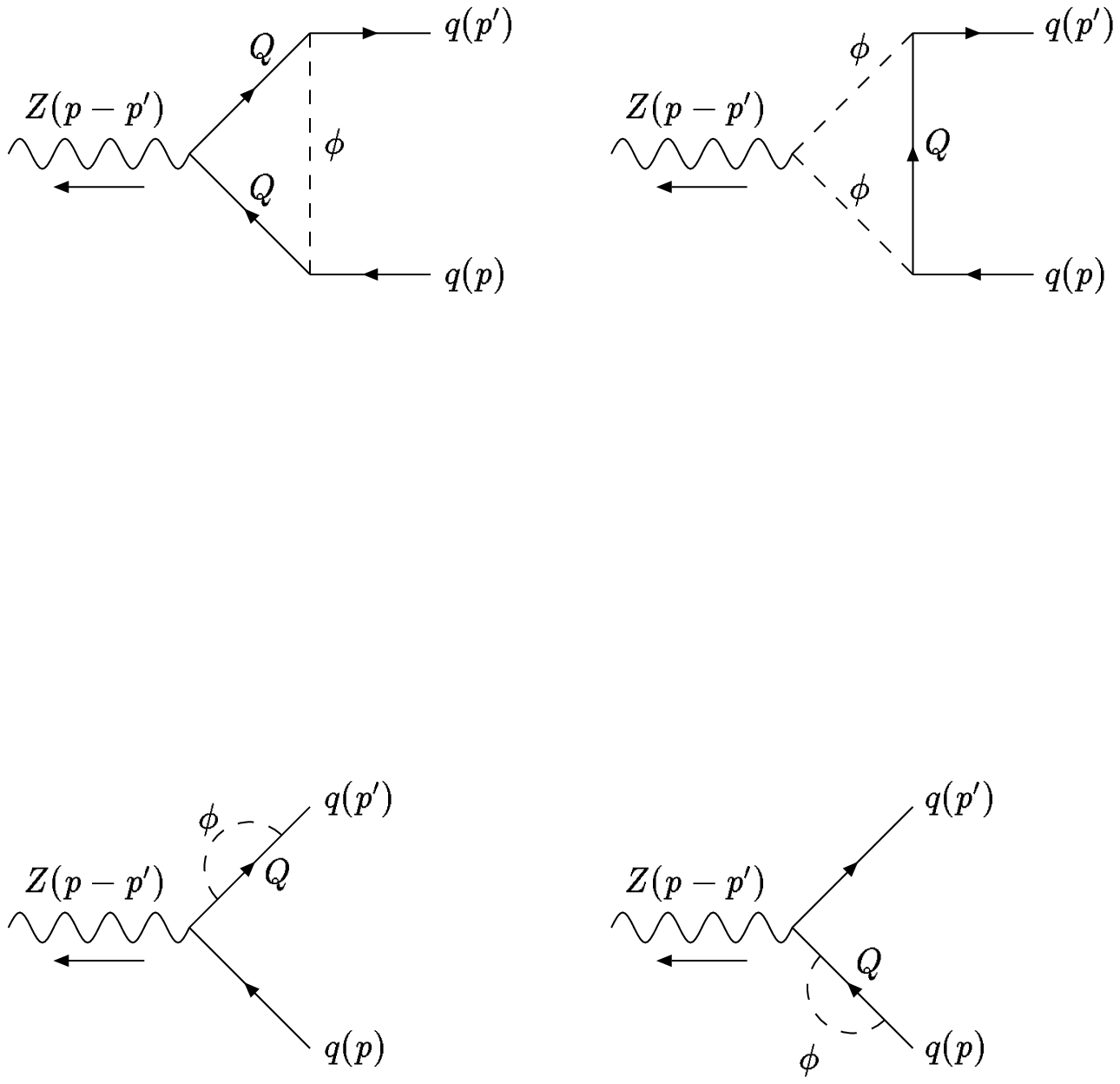}}

\vspace{-20ex}
\caption{Typical diquark--induced diagrams leading to additional
contribution to \protect $\Gamma(Z \rightarrow q \bar{q})$.}
       \label{diag}
\end{figure}

\begin{figure}[htb]
\vskip 8in\relax\noindent\hskip -1in\relax{\includegraphics{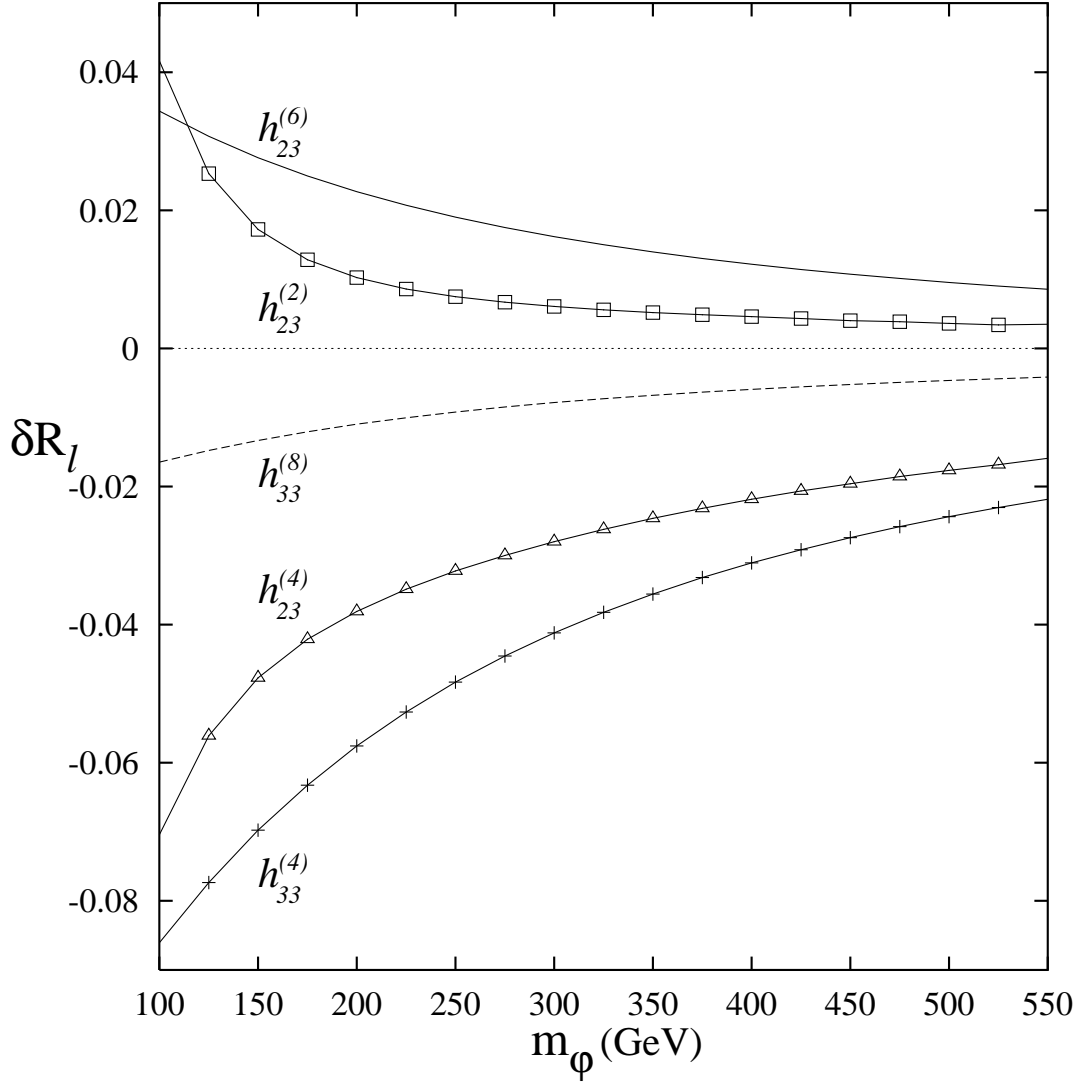}}

\vspace{-20ex}
\caption{The contribution to $R_l$ in the presence of different
$B$-violating couplings with  $SU(3)_c$-triplet scalars.
For each individual curve, the concerned Yukawa coupling
has been assumed to be unity
while all other $B$-violating couplings are held to be zero.}
\label{rl}
\end{figure}

\begin{figure}[htb]
\vskip 8in\relax\noindent\hskip -1in\relax{\includegraphics{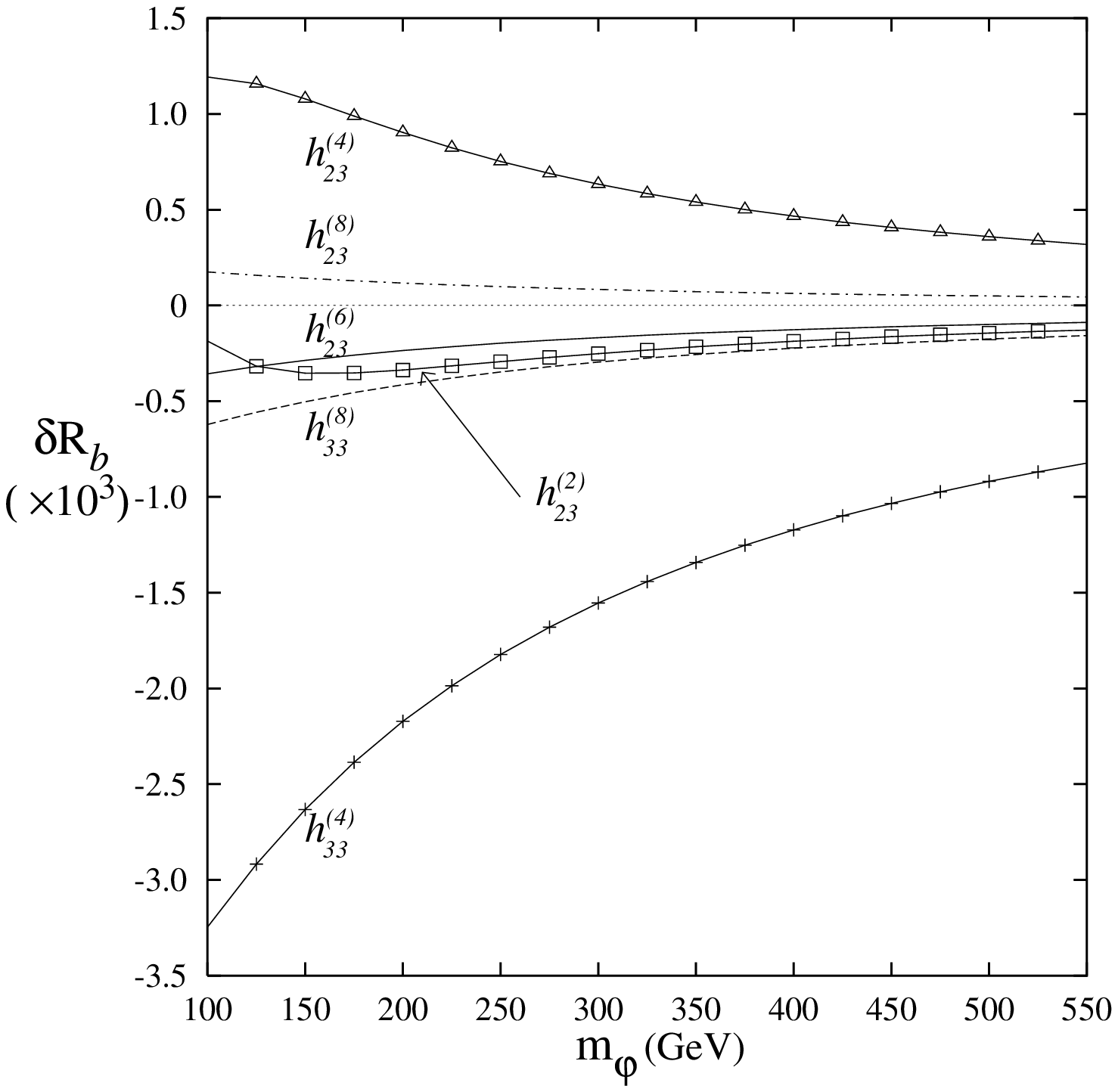}}

\vspace{-20ex}
\caption{As in Fig. \protect\ref{rl}, but for $R_b$}
       \label{rb}
\end{figure}

\begin{figure}[htb]
\vskip 8in\relax\noindent\hskip -1in\relax{\includegraphics{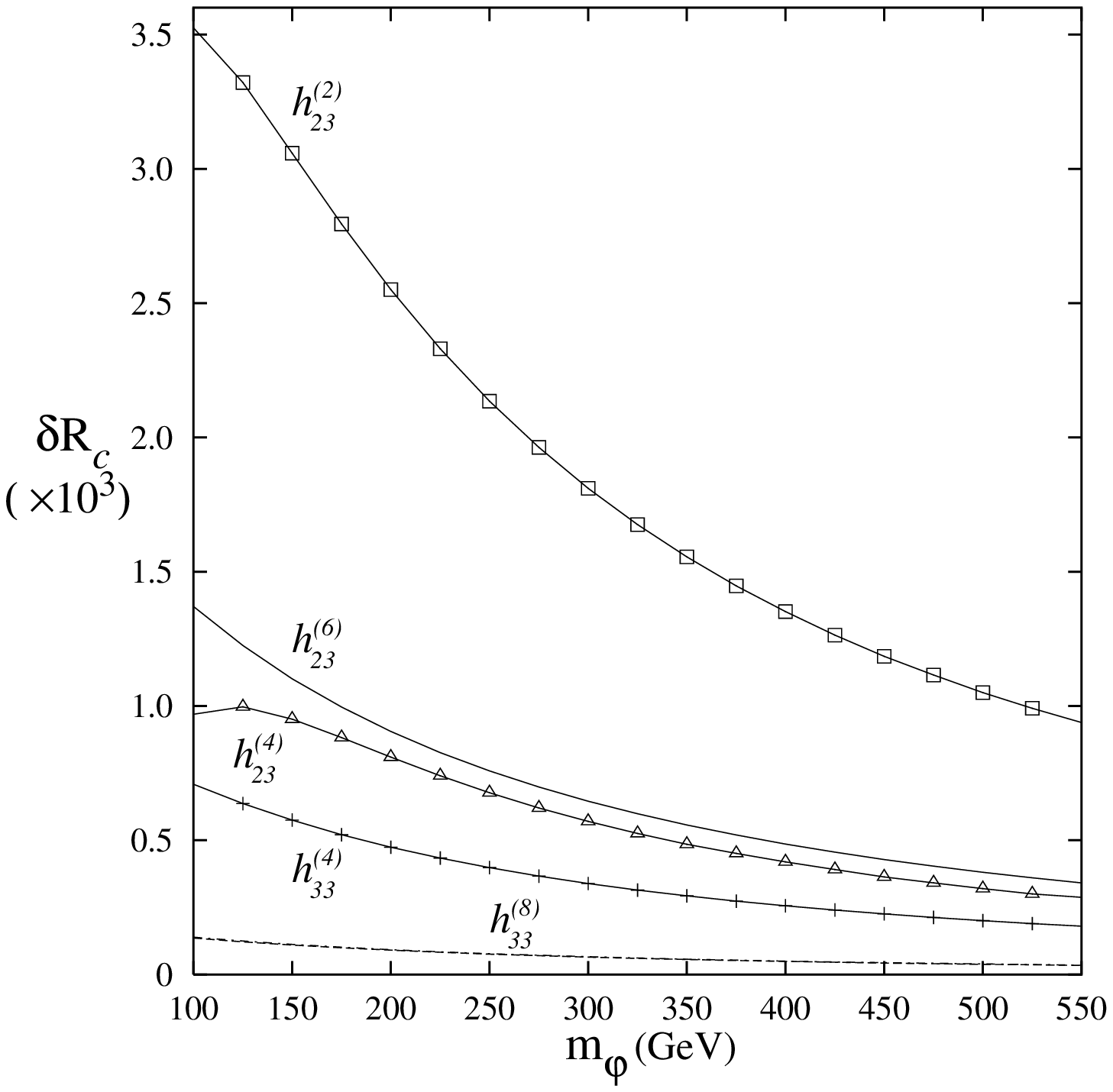}}

\vspace{-20ex}
\caption{As in Fig.\protect\ref{rl}, but for $R_c$}
       \label{rc}
\end{figure}

\end{document}